\documentclass[times]{cpeauth}

\usepackage{moreverb}

\usepackage{doc}
\usepackage{makeidx}

\newfont{\toto}{msbm10 at 12 pt}

\newcommand{\mc}[3]{\multicolumn{#1}{#2}{#3}}
\newcommand{\cc}[2]{\multicolumn{1}{#1}{#2}}

\newcommand{\bc}{\begin{center}}
\newcommand{\ec}{\end{center}}

\newcommand{\bd}{\begin{description}}
\newcommand{\ed}{\end{description}}

\newcommand{\bi}{\begin{itemize}}
\newcommand{\ei}{\end{itemize}}

\newcommand{\benu}{\begin{enumerate}}
\newcommand{\eenu}{\end{enumerate}}

\newcommand{\bq}{\begin{quote}}
\newcommand{\eq}{\end{quote}}

\newcommand{\be}{\begin{equation}}
\newcommand{\ee}{\end{equation}}
\newcommand{\bea}{\begin{eqnarray}}
\newcommand{\eea}{\end{eqnarray}}

\newcommand{\T}{T}
\newcommand{\dt}{{\Delta t}}

\usepackage{graphicx}
\usepackage{color}
\usepackage{listings}
\usepackage{bm}
\usepackage{url}
\usepackage{booktabs}
\usepackage{amsmath}
\usepackage{paralist}
\usepackage{todonotes}
\usepackage{booktabs}

\definecolor{g90}{gray}{.90}
\definecolor{myBlue}{rgb}{0,0,.59}
\definecolor{myGreen}{rgb}{0,.20,.2}
\definecolor{myBrown}{rgb}{.59,0,0}

\usepackage{amssymb}

\begin{document}


\runningheads{E.~Calore at al.}{Performances and Portability of 
Accelerated LB Applications with OpenACC}

\title{Performance and Portability of Accelerated Lattice Boltzmann Applications with OpenACC}

\author{
E.~Calore\affil{1}, 
A.~Gabbana\affil{1},
J.~ Kraus\affil{2}, 
S.~F.~Schifano\affil{3}\corrauth,
R.~Tripiccione\affil{1}
}

\address{
\affilnum{1}Dip. di Fisica e Scienze della Terra, University of Ferrara, and INFN, Ferrara (Italy)\break
\affilnum{2}NVIDIA GmbH, W\"urselen (Germany)\break
\affilnum{3}Dip. di Matematica e Informatica, University of Ferrara, and INFN, Ferrara (Italy)
}

\corraddr{
via Saragat 1, I-44124 FERRARA (Italy). 
Email: schifano@fe.infn.it. 
Tel/Fax: +390532974614
}


\begin{abstract}

An increasingly large number of HPC systems rely on heterogeneous 
architectures combining traditional multi-core CPUs with power 
efficient accelerators.
Designing efficient applications for these systems has been troublesome 
in the past as accelerators could usually be programmed using specific 
programming languages threatening maintainability, portability and 
correctness.
Several new programming environments try to tackle this problem. 
Among them, OpenACC offers a high-level approach based on compiler 
directives to mark regions of existing C, C++ or Fortran codes 
to run on accelerators.
This approach directly addresses code portability, leaving to compilers 
the support of each different accelerator, but one has to carefully 
assess the relative costs of portable approaches versus computing efficiency. 
In this paper we address precisely this issue, using as a test-bench 
a massively parallel Lattice Boltzmann algorithm. We first describe 
our multi-node implementation and optimization of the algorithm, 
using OpenACC and MPI. 
We then benchmark the code on a variety of processors, including 
traditional CPUs and GPUs, and make accurate performance comparisons 
with other GPU implementations of the same algorithm using CUDA 
and OpenCL. We also asses the performance impact associated to portable 
programming, and the actual portability and performance-portability 
of OpenACC-based applications across several state-of-the-art architectures.

\textbf{This is the pre-peer reviewed version of the following article: 
Performance and Portability of Accelerated Lattice Boltzmann Applications with OpenACC. (2016)
Concurrency Computat.: Pract. Exper., 28: 3485-3502, 
which has been published in final form at 10.1002/cpe.3862. 
This article may be used for non-commercial purposes in accordance with Wiley Terms and Conditions for Self-Archiving.}

\end{abstract}



\maketitle


\section{Introduction, Related Works and Background}

Lattice Boltzmann (LB) methods are widely used in computational fluid 
dynamics, to simulate flows in two and three dimensions. 
From the computational point of view, LB methods have a large degree of
available parallelism so they are suitable for massively parallel systems. 

Over the years, LB codes have been written and optimized for large clusters 
of commodity CPUs~\cite{Pohl2004}, for application-specific machines
~\cite{cimento09,iccs10,ppam13} and even for FPGAs~\cite{lbm-fpga}.
More recently work has focused on exploiting the parallelism of powerful 
traditional many-core processors~\cite{iccs11}, and of power-efficient 
accelerators such as GPUs~\cite{ppam11,lbm-gpu2} or Xeon-Phi processors~\cite{iccs13}.
  
As diversified HPC architectures emerge, it is becoming more and more important
to have robust methodologies to port and maintain codes for several
architectures. This need has sparked the development of frameworks, such as the
{\em Open Computing Language} (OpenCL), allowing to write portable codes, 
that can be compiled (with varying degrees of efficiency) for 
several accelerator architectures.
OpenCL is a low level approach: it usually obtains high performances at the 
price of substantial adjustments in the code implying large human efforts and 
seriously posing a threat to code correctness and maintainability.

Other approaches start to emerge, mainly based on directives: compilers
generate offload-functions for accelerators, following ``hints'' provided by
programmers as  annotations to the original -- C, C++ or Fortran -- codes~\cite{first-exp}.
Examples along this direction are OpenACC~\cite{openacc} and 
OpenMP4~\cite{openmp4}.
Other proposals, such as the Hybrid Multi-core Parallel Programming 
model (HMPP) proposed by CAPS, hiCUDA~\cite{hicuda}, OpenMPC~\cite{openmpc} 
and StarSs~\cite{starss} follow the same line.

OpenACC today is considered among the most promising approaches to develop 
high-performance scientific applications~\cite{se4hpcs15}. 
In many ways its structure is similar to OpenMP (Open Multi-Processing)~\cite{Wienke2014812}:
both frameworks are directive based, but while OpenMP is more {\em prescriptive} 
OpenACC is more {\em descriptive}. 
Indeed, with OpenACC the programmer only specifies that a certain loop should 
run in parallel on the accelerator and leaves the exact mapping to the compiler.
This approach gives more freedom to the compiler and the associated runtime
support, offering -- at least in principle -- larger scope for performance portability.

So far very few OpenACC implementations of LB codes have been described in
literature: \cite{jiri} focuses on accelerating via OpenACC a part of a large 
CFD application optimized for CPU; several other works describe CUDA~\cite{sbac-pad13} 
or OpenCL~\cite{iccs14} implementations; also the scalability of OpenACC codes on GPU 
clusters has been rarely addressed~\cite{scalable-lbm-cuda}. 
This work describes the implementation of a state-of-the-art LB code fully 
written in OpenACC, including accurate performance measurements and an 
assessment of the actual portability improvements made possible by this 
programming style.
This is an extended version of~\cite{europar15}, that we have presented 
at the {\em EuroPar 2015} conference.  
In the original paper, we focused on the design and optimization of a multi-GPU LB 
code, discussing performance trade-offs between a portable approach based on 
OpenACC and processor-specific languages such as CUDA.
The present work includes additional material, discussing and analyzing issues 
related to the portability as well as the performance portability of our OpenACC 
codes. 
Our analysis is based on tests performed on several 
computing architectures, including multi-core CPUs and several different GPUs.
In other words, the original paper provided an answer to the question of the 
performance price that one has to pay if one uses  OpenACC instead of a 
processor-specific programming language,  while this extended version also 
answers the question of how portable and {\em performance-}portable 
is just one architecture-oblivious OpenACC code across a fairly large set of 
different architectures.

Very recently {\em Blair et al.} have described an implementation of a 
MPI Lattice Boltzmann code with OpenACC~\cite{blair15}; however portability of code and 
performances across different architectures have not been analyzed;  
to the best of our knowledge, this paper is the first work discussing these 
issues for OpenACC.

This paper is structured as follows: Sect.~\ref{sec:lbm} gives a short 
overview of LB methods and Sect.~\ref{sec:openacc} a quick overview of the
OpenACC programming framework;
Sect.~\ref{sec:implementation} describes in details our OpenACC implementation, 
and Sect.~\ref{sec:results} analyzes performance results on GPUs in comparison with a
CUDA implementation of the same code.
Section~\ref{sec:portability} (added in the extended version) analyzes 
the portability of the {\em same} OpenACC code on different architectures:
Intel E5-2630 v3 multicore CPUs, NVIDIA K80 and AMD S9150 GPUs. 
Finally, Sect.~\ref{sec:conclusions} highlights our conclusions.


\section{Lattice Boltzmann Models}\label{sec:lbm}

Lattice Boltzmann methods (LB) are widely used in computational fluid dynamics,
to describe flows in two and three dimensions. LB methods~\cite{sauro} are discrete 
in position and momentum spaces; they are based on the synthetic dynamics of {\em populations}
sitting at the sites of a discrete lattice. At each time step, populations hop
from lattice-site to lattice-site and then incoming populations {\tt collide}
among one another, that is, they mix and their values change accordingly. 

Over the years, many different LB models have been developed, handling 
flows in 2 and 3  dimensions with different degrees of accuracy~\cite{aidun}. 
LB models in $n$ dimensions with $y$ populations are labeled as $DnQy$;  
in this paper, we consider a state-of-the-art $D2Q37$ model that correctly reproduces the
thermo-hydrodynamical equations of motion of a fluid in two dimensions and
automatically enforces the equation of state of a perfect gas ($p = \rho T$)
~\cite{JFM,POF}; this model has been extensively used for large scale 
simulations of convective turbulence (see e.g.,~\cite{noi1,noi2,ripesi14}).
 
In our model, populations 
($f_l({\bm x},t)~l = 1 \cdots 37$) are defined at the sites of a discrete and 
regular 2-D lattice; each $f_l({\bm x},t)$ has a given lattice velocity $\bm c_{l}$;
populations evolve in (discrete) time according to the following equation
(the BGK operator~\cite{bgk}):
\begin{equation}
f_{l}({\bm x}, t+\dt) = f_{l}({\bm x} - {\bm c}_{l} \dt,t) 
-\frac{\dt}{\tau}\left(f_{l}({\bm x} - {\bm c}_{l} \dt,t) - f_l^{(eq)}\right)
\label{eq:master2}
\end{equation}

Macroscopic quantities, density $\rho$, velocity $\bm u$  and temperature $T$
are defined in terms of the $f_l(x,t)$ and of the $\bm c_{l}$s ($D$ is the 
number of space dimensions):
\begin{eqnarray}
\rho = \sum_l f_l,~~~
\rho {\bm u} = \sum_l {\bm c}_l f_l, ~~~ 
D \rho \T = \sum_l \left|{\bm c}_l - {\bm u}\right|^2 f_l;
\label{eq:macro}
\end{eqnarray}
the equilibrium distributions (${f}_{l}^{(eq)}$) are known functions of
these macroscopic quantities \cite{sauro}, and $\tau$ is a suitably chosen relaxation time. 
In words, (\ref{eq:master2}) stipulates that populations drift from lattice
site to lattice site according to the value of their velocities 
({\em propagation}) and, on arrival at point ${\bm x}$, they interact among  one another and
their values change accordingly ({\em collision}). One can show that, in
suitable limiting cases and after appropriate renormalizations are applied, the
evolution of the macroscopic variables defined in (\ref{eq:macro}) obeys the
thermo-hydrodynamical equations of motion of the fluid.

From a computational point of view the physically very accurate LB scheme that we adopt  is
more complex than many simpler LB models. One specific optimization step of the algorithm applies a systematic projection onto a basis of Hermite polynomials to improve numerical stability; this translates into severe 
requirements in terms of required storage, memory bandwidth and floating-point throughput (at each time step, $\approx 7600$ 
double-precision floating point operations are performed per lattice point).

%
\begin{figure}[t]
\centering
\begin{minipage}[h]{0.4\textwidth}
\includegraphics[width=\textwidth]{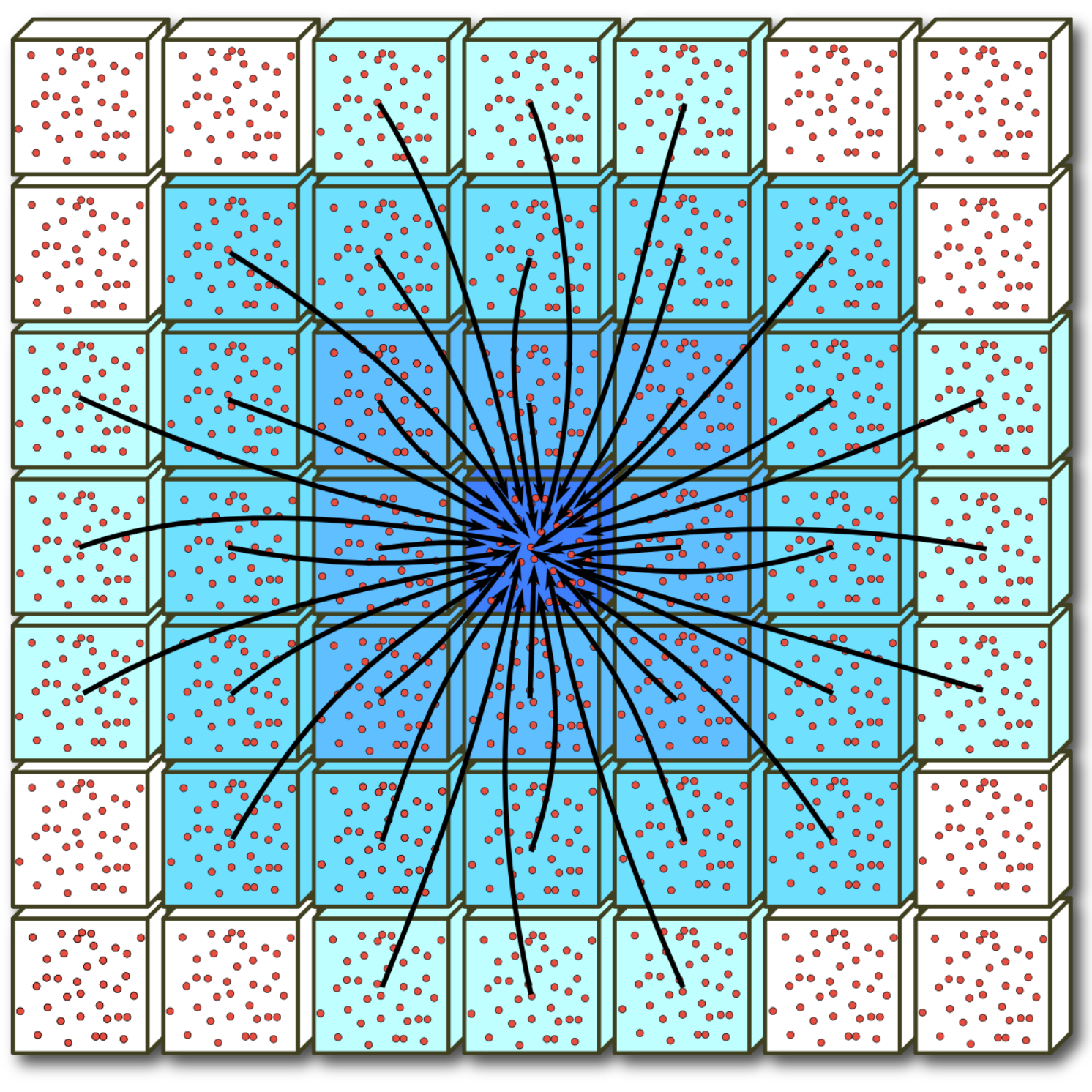}
\end{minipage}
\hspace*{20mm}
\begin{minipage}[h]{0.4\textwidth}
\includegraphics[width=\textwidth]{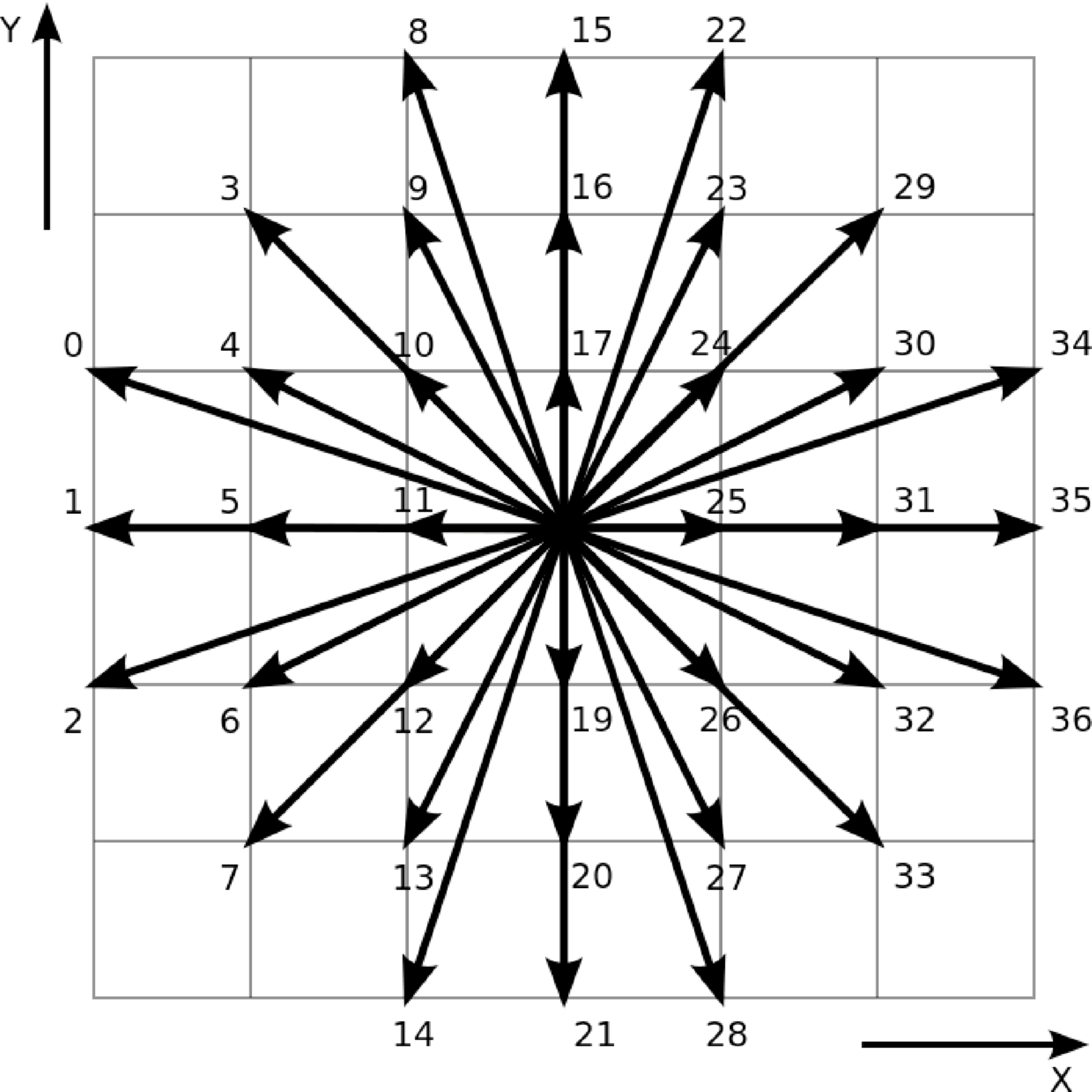}
\end{minipage}
\caption{Left: LB populations in the D2Q37 model, hopping to nearby 
sites during the {\tt propagate} phase.
Right: populations $f_l$ are identified by an arbitrary label; 
for each $l$ population data is stored contiguously in memory.}
\label{streamscheme}
\end{figure}
%

An LB code takes an initial assignment of the populations, in accordance
with a given initial condition at $t = 0$ on some spatial domain, and iterates
(\ref{eq:master2}) for all points in the domain and for as many time-steps
as needed; boundary-conditions at the edges of the integration
domain are enforced at each time-step by appropriately modifying population 
values at and close to the boundaries.

The LB approach offers a huge degree of easily identified parallelism. Indeed,
(\ref{eq:master2}) shows that the {\em propagation} step amounts
to gathering the values of the fields $f_l$ from neighboring sites,
corresponding to populations drifting towards ${\bm x}$ with velocity 
${\bm c}_l$; the following step ({\em collision}) then performs all mathematical
processing needed to compute the quantities in the r.h.s. of 
(\ref{eq:master2}), for each point in the grid. One sees immediately from 
(\ref{eq:master2}), that both steps above are fully
uncorrelated for different points of the grid, so they can be executed in 
parallel according to any schedule, as long as step 1 precedes step 2
for all lattice points.

In practice, an LB code executes a loop over time steps, and at each 
iterations applies three kernels: {\tt propagate}, {\tt bc} and {\tt collide}.

\begin{itemize}
\item
{\tt propagate} moves populations across lattice sites according to the
pattern of~\figurename~\ref{streamscheme}, collecting at each site
all populations that will interact at the next phase ({\tt collide}).
In our model populations move up to three lattice sites per time step.
Computer-wise, {\tt propagate} moves blocks of  memory locations allocated
at sparse addresses, corresponding to populations of neighbor cells. 

\item 
{\tt bc} executes {\em after} propagation and adjusts populations at the
edges of the lattice, enforcing appropriate boundary conditions 
(e.g., constant temperature and zero velocity at 
the top and bottom edges of the lattice).
For the left and right edges, we usually apply periodic 
boundary conditions. 
This is conveniently done by adding {\em halo} columns at the
edges of the lattice, where we copy the rightmost and leftmost 
columns (3 in our case) of the lattice before starting
the {\tt propagate} step.
After this is done, points close to the boundaries are processed 
as those in the bulk.  

\item
{\tt collide} performs all mathematical steps needed to compute the population  
values at each lattice site at the new time step, as per (\ref{eq:master2}). 
Input data for this phase are the populations gathered by the previous 
{\tt propagate} phase. This step is the most floating point intensive 
part of the code.
\end{itemize}

These three routines use essentially all the wall-clock time of a typical 
LB production run, as additional measurement routines, typically computing
averages or correlations of physical observables, are invoked once 
every several hundreds or even thousands of time steps. 

%
\begin{figure}
\centering
\definecolor{mygray}{rgb}{0.5,0.5,0.5}
\lstset{escapeinside={(*@}{@*)}}
\begin{lstlisting}[basicstyle=\footnotesize,language=C,numbers=left,
numberstyle=\color{mygray},stepnumber=1,numbersep=-3.75pt]
 #pragma acc copyin(x), copyout(y) (*@\label{lst:data}@*)
 {
    // asynchronous kernel execution 
    #pragma acc kernels present(x) present(y) async(1) (*@\label{lst:kernels}@*)
    {
      #pragma acc loop 
      for (int i = 0; i < N; ++i) (*@\label{lst:init-loop}@*)
        y[i] = 0.0;
      #pragma acc loop device_type(NVIDIA) gang vector(256) (*@\label{lst:gang-vector}@*)
      for (int i = 0; i < N; ++i) (*@\label{lst:saxpy-loop}@*)
        y[i] = a*x[i] + y[i];
    }
   
   //... process independent data on the host
   
   // wait for completion of kernel execution
   #pragma acc wait(1) (*@\label{lst:wait}@*)
   
   // ... other processing on the host
 }
\end{lstlisting}
\caption{\label{lst:saxpy} Sample OpenACC code computing a {\em saxpy}
function on vectors $x$ and $y$. Directives mark the 
code region to run on the accelerator and instruct the compiler on how to 
generate code for the target device.}
\end{figure}
%

\section{OpenACC}\label{sec:openacc}

OpenACC is a programming standard for parallel computing aimed to 
facilitate code development on heterogeneous computing systems,
simplifying the porting of existing codes and trying to achieve a 
significant level of performance portability.

Its support for different architectures relies on compilers, and thanks 
to its generality the same code can be compiled and parallelized for different 
target architectures if the corresponding back-end and run-time support are available. 
Recent versions of OpenACC implementations, such as the PGI release, version 15.10, 
can address as target architectures NVIDIA and AMD accelerators as well as 
commodity multicore {\em x}86 CPUs~\cite{nvidia-portability}.

OpenACC, like OpenCL, provides a widely applicable abstraction of parallel 
hardware, making it possible to run the same code across different 
architectures.
Contrary to OpenCL, where specific functions (called \textit{kernels}) have to
be explicitly programmed to run in a parallel fashion (e.g. as GPU threads), 
OpenACC relies on developer-provided directives that help the compiler identify those 
parts of the source code that can be implemented as {\em parallel functions}.
Following these {\em directives} the compiler generates one or more 
\textit{kernel} functions -- in the OpenCL sense -- to be executed in parallel 
by many threads.

OpenACC is similar to the OpenMP (Open Multi-Processing) language in several 
ways~\cite{Wienke2014812}; both environments are directive based, but OpenACC 
targets accelerators in general, while so far OpenMP has been used to target 
mainly multi-core CPUs. The latest version of OpenMP~\cite{openmp4} has recently 
introduced support for accelerators using a model very close to that of 
OpenACC. From a programmers perspective, OpenMP is more prescriptive in the 
sense that explicit mapping of work-loads to compute-units, e.g. using distribute 
constructs, is required. In contrast to this OpenACC, being more descriptive,
only requires the programmer to expose parallelism and let the compiler do the 
actual mapping to the compute-units.
Since different hardware architectures require different mappings to perform 
efficiently, this approach makes OpenACC in principle more performance portable. 

%
\begin{figure}[t]
\centering
\begin{lstlisting}[language=C,basicstyle=\footnotesize]
// processing of lattice-bulk
propagateBulk( f2, f1 ); // async execution on queue (1)
bcBulk( f2, f1 );        // async execution on queue (1)         
collideInBulk( f2, f1 ); // async execution on queue (1)

// execution of pbc step
#pragma acc host_data use_device(f2) {
  for ( pp = 0; pp < 37; pp++ ) {
    MPI_Sendrecv ( &(f2[...]), 3*NY, ... );
    MPI_Sendrecv ( &(f2[...]), 3*NY, ... );
  } 
}

// processing of the three leftmost columns
propagateL( f2, f1 );   // async execution on queue (2)
bcL( f2, f1 );          // async execution on queue (2)  
collideL( f1, f2 );     // async execution on queue (2)

// processing of the three rightmost columns
propagateR( f2, f1 );   // async execution on queue (3)
bcR( f2, f1  );         // async execution on queue (3)  
collideR( f1, f2  );    // async execution on queue (3)
\end{lstlisting}
\caption{Scheduling of operations started by the host at each time 
step of the main program. Kernels processing the lattice bulk run  
asynchronously on the accelerator, and overlap with MPI communications executed by the host.}
\label{fig:schedulingop}
\end{figure}
%

Existing C/C++ or Fortran code, developed and tested on traditional
CPU architectures, can be annotated with OpenACC directives (e.g. 
\textit{parallel} or \textit{kernels} directives) to instruct the compiler to 
transform loop iterations into distinct threads, belonging to one or more 
functions to run on an accelerator. \figurename~\ref{lst:saxpy} shows a simple example
based on the \textit{saxpy} operation from the {\em Basic Linear Algebra Subprogram} (BLAS) set.

Line~\ref{lst:kernels} contains the \textit{pragma acc kernels} directive 
which identifies the code to run on the accelerator; in this case the 
iterations of the two for-loops are parallelized and the execution of the function is offloaded 
at run-time from the host CPU to an attached accelerator device (e.g. a GPU). 
More directives are available, allowing a finer tuning of the application.
As an example, the number of threads launched by each device function and
their grouping can be fine tuned by the \textit{vector}, \textit{worker} and
\textit{gang} directives, in a similar fashion as setting the number of
\textit{work-items} and \textit{work-groups} in OpenCL: in the example of
\figurename~\ref{lst:saxpy}, line~\ref{lst:gang-vector} sets vector length to 256.
Since the \textit{gang} and \textit{vector} clauses encode a hardware specific tuning
they follow a \textit{device\_type} clause, e.g., they only apply to the
target specified as an argument of the \textit{device\_type} clause and allow 
hardware specific tuning without harming performance portability.

Data transfers between host and device memories are automatically generated,
when needed. These automatically generated data movements are often overly
cautious and thus can be optimized by the programmer with data directives. 
For example, in the code shown in \figurename~\ref{lst:saxpy} the clause 
\textit{copyin(x)} (at line~\ref{lst:data}) copies the array of the host CPU 
pointed by \textit{x} onto the accelerator memory before entering the
following code region; while \textit{copyout(y)} allocates an empty buffer
before entering the region and copies it back to the host memory after leaving.
Introducing this so called data region avoids multiple unnecessary copies:
\begin{enumerate}
\item a copy of \textit{y} from host memory to the accelerator memory before 
the loop in line~\ref{lst:init-loop} and the reverse copy after the loop.
\item a copy of \textit{y} from host memory to the accelerator memory before 
the loop in line~\ref{lst:saxpy-loop}.
\item a copy of \textit{x} from the accelerator memory to host memory after 
the loop in line~\ref{lst:saxpy-loop}.
\end{enumerate}

The last OpenACC feature used in the example is the asynchronous clause \textit{async}
(at line~\ref{lst:kernels}) which instructs the compiler to generate asynchronous data
transfers or kernel executions and allows overlapping of independent data transfers,
kernels and CPU work. A directive corresponding to the \textit{async} clause is
provided by the OpenACC API (\textit{\#pragma acc wait} at line~\ref{lst:wait})
which allows to wait for completion. 
For more details on OpenACC features and functions see~\cite{openacc}.

%
\begin{figure}[t]
\centering
\begin{lstlisting}[language=C,basicstyle=\footnotesize]
inline void propagate ( 
  const data_t* restrict prv, data_t* restrict nxt ) {
  int ix, iy, site_i;
  
  #pragma acc kernels present(prv) present(nxt)
  #pragma acc loop gang independent
  for ( ix=HX; ix < (HX+SIZEX); ix++) {
    #pragma acc loop vector independent
    for ( iy=HY; iy<(HY+SIZEY); iy++) {
      site_i = (ix*NY) + iy;
      nxt[      site_i] = prv[      site_i-3*NY+1];
      nxt[NX*NY+site_i] = prv[NX*NY+site_i-3*NY  ];
      ....
    } 
  } 
  
}
\end{lstlisting}
\caption{\label{fig:propagate} OpenACC pragmas in the body of the 
{\tt propagate()} function; pragmas before the loops instruct the 
compiler to generate corresponding accelerator 
kernels and to configure the grid of threads and blocks.}
\end{figure}
%


\section{OpenACC Implementation and Optimization of the D2Q37 Model}\label{sec:implementation}

This section describes in details the strategies that we have adopted 
to write an OpenACC version of our LB code suitable for compilation 
and execution on NVIDIA GPUs. In a later section, we will then focus 
on portability issues, as we experiment with this code on different HPC 
architectures and measure the corresponding performances.

One of our initial goals was to have a massively parallel program, 
able to run on a large number of GPUs.
From the point of view of data organization, we adopt a very simple domain 
decomposition, splitting our 2-D physical lattice of size $L^{tot}_x \times L_y$ 
on $N$ accelerators along the $X$ dimension; GPUs are connected in a ring-scheme,
each hosting a sub-lattice of $L^{tot}_x/N \times L_y = L_x \times L_y$ points. 
We use MPI for the overall control of node-parallelism, starting one MPI rank
for each GPU, so  GPU-to-GPU transfers are transparently handled by the MPI
library; once this is done, we use OpenACC to annotate the code executed by each
MPI rank. 

On each MPI-rank the physical lattice is surrounded by halo columns and rows: 
for a physical sub-lattice of size $L_x \times L_y$, we allocate $NX \times NY$ 
points, with $NX=H_x+L_x+H_x$ and $NY=H_y+L_y+H_y$.
With this splitting, halo-columns are allocated at successive 
memory locations, so we do not need to gather halo data on 
contiguous buffers before communication. 

Data is stored in memory in the Structure-of-Array (SoA) scheme,
where arrays of all populations are stored one after the other. This helps
exploit data-parallelism and enables data-coalescing when accessing data
needed by work-items executing in parallel.

The lattice is copied on the accelerator memory at the beginning of the 
loop over time-steps, and then all three kernels of the algorithm 
-- {\tt propagate}, {\tt bc} and {\tt collide} -- run in sequence 
on the accelerator for as many time-steps as needed.
Merging in one step execution of {\tt propagate} and 
{\tt collide} is a common optimization in LBM codes. 
However, in our case this would require a more complex organization 
of the code to process separately lattice-sites that have no dependencies with 
{\tt bc} kernel that runs before {\tt collide} but after {\tt propagate}. 
For this reason our implementation runs the two kernels in two separate steps; 
this is also useful for benchmark purposes since these kernels have different 
computing  requirements being -- in our LB model -- the first memory-bound 
and the latter strongly compute-bound    

The execution of these kernels starts after an update of the left- and 
right-halos is performed:
we copy population data coming from the three adjoining physical columns 
of the neighbor nodes in the ring to the left and right {\em halos}.
%
This is done by an MPI node-to-node communication step that we call 
{\em periodic boundary condition} ({\tt pbc}).
Once this is done, all remaining steps are local to each MPI-rank so 
they run in parallel.

As lattice data is stored in the SoA format, {\tt pbc} exchanges 37
buffers, each of 3 columns, with its left  and right neighbors. 
It executes a loop over the 37 populations and each iteration performs 
two MPI send-receive operations, respectively for the left and the 
right halo (see \figurename~\ref{fig:schedulingop}).
On GPUs, we exploit {\em CUDA-aware} MPI features, available in the OpenMPI 
library, and use data pointers referencing GPU-memory 
buffers as source and destination, making the code more compact and readable.
In OpenACC this is controlled by the {\tt \#pragma acc host\_data
use\_device(p)} clause, that maps a GPU memory pointer {\tt p} into host
space, so it can be used as an argument of the MPI send and receive functions.
Also, communications between GPUs are optimized in the library 
and implemented according to physical location of buffers and the 
capabilities of the devices involved, also enabling {\em GPUDirect} 
peer-to-peer and {\em RDMA} features. 

%
\begin{figure}
\centering
\includegraphics[width=\textwidth]{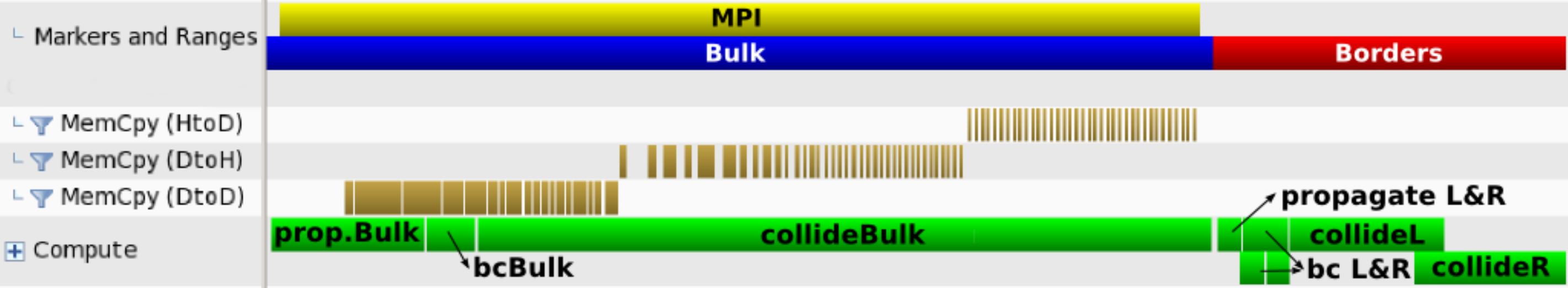}
\caption{\label{fig:scheduling} 
Profiling of one time step. In this example,
{\tt pbc} (yellow line marked as ``MPI'') and the kernels processing 
the bulk of the lattice (blue line marked as ``Bulk'') fully overlap.
}
\end{figure}
%

Coming now to the main kernels of the algorithm, \figurename~\ref{fig:propagate} 
shows the code of the {\tt propagate} function. 
For each lattice site we update the values of the populations, 
copying from the {\tt prv} array onto the {\tt nxt} array. 
The body of {\tt propagate} is annotated with several OpenACC 
directives telling the compiler how to organize the kernel
on the accelerator. 
{\tt \#pragma acc kernels present(prv) present(nxt)} tells the compiler to run the
following instructions on the accelerator; it also carries the information
that the {\tt prv} and {\tt nxt} arrays are already available on the accelerator
memory, so no host-accelerator data transfer is needed;
{\tt \#pragma acc loop gang independent} states that each iteration 
of the following loop (over the X-dimension) can be run by different gangs or
block of threads;
{\tt \#pragma acc loop vector independent} tells the compiler that  
iterations of the loop over Y-dimension can likewise be run as 
independent vectors of threads. 
Using these directives the compiler structures the thread-blocks and 
block-grids of the accelerator computation in such a way that: one thread is associated 
to and processes one lattice-site; each thread-block processes a group 
of lattice sites lying along the Y-direction, and several blocks 
process sites along the X-direction. This allows to expose all available 
parallelism. 

We split {\tt bc()} in two kernels,  
processing the upper and lower boundaries. They run in parallel 
since there is no data dependencies among them.
We have not further optimized this step because its computational cost 
is small compared to the other phases of the code.

The {\tt collide()} kernel sweeps all lattice sites and computes 
the collisional function. The code has two outer loops over the two  
dimensions of the lattice, and several inner loops to compute temporary 
values. We have annotated the outer loops as we did for {\tt propagate()}, 
making each thread to process one lattice site. Inner 
loops are computed serially by the thread associated to each site.


Performance wise, {\tt pbc()} is the most critical step of the multi-GPU 
code, since it involves node-to-node communications that can badly affect 
performance and scaling.
We organize the code so 
node-to-node communications are (fully or partially) overlapped  
with the execution of other segments of the code. 
Generally speaking, {\tt propagate}, {\tt bc} and  {\tt collide} must execute
one after the other, and they cannot start  before {\tt pbc} has completed. 
One easily sees however that this dependency does not apply to all sites of the
lattice outside the three leftmost and rightmost border columns
(we call this region the {\em bulk} of the lattice). 
The obvious conclusion is that processing of the bulk can proceed in parallel 
with the execution of {\tt pbc}, while the sites on the three leftmost and 
rightmost columns are processed only after {\tt pbc} has completed.

OpenACC abstracts concurrent execution using queues: function 
definitions flagged by the {\tt \#pragma acc async(n)} directive enqueue 
the corresponding kernels {\em asynchronously} on queue {\tt n}, 
leaving the host free to perform other tasks concurrently.
In our case, this happens for {\tt propagateBulk}, 
{\tt bcBulk} and {\tt collideBulk}, which start on queue 1 (see~\figurename~\ref{fig:schedulingop}), 
while the host concurrently executes the MPI transfers of {\tt pbc}.
After communications complete, the host starts three more kernels 
on two different queues (2 and 3) to process the right and left borders, 
so they execute in parallel if sufficient resources on the accelerator are available.
This structure allows to overlap {\tt pbc} with  
all other steps of the code, most importantly with {\tt collideBulk}, which is
the most time consuming kernel, giving more opportunities to hide 
communication overheads when running on a large number of nodes.

\figurename~\ref{fig:scheduling} shows the profiling of one time step  
on one GPU on a lattice of $1080 \times 2048$ points split across 24 GPUs.
MPI communications started by {\tt pbc} are internal (MemCopy DtoD),
moving data between GPUs on the same host, or external (MemCopy DtoH and HtoD) 
moving data between GPUs on different hosts.
The actual scheduling is as expected: both types of GPU-to-GPU communications 
fully overlap with {\tt propagate}, {\tt bc} and {\tt collide} on the bulk.

%
\begin{table}[t]
\caption{
Performance comparison of OpenACC code with CUDA version running on 
NVIDIA Tesla K40 and K80 GPU accelerator cards; all codes run on a 
lattice size of $1920 \times 2048$ points. 
All quantities are defined in the text. The last two rows show the 
``wall-clock'' execution time and the corresponding MLUPS 
({\em Millions Lattice UPdate per Second}) for the full code.
}
\label{comparison}
\centering
\begin{tabular}{l rrr rr rrr rr}
\toprule
                             &&&&  \mc{2}{c}{Tesla K40} 	    &&&& \mc{2}{c}{Tesla K80}	   \\
\cmidrule{5-6} \cmidrule{10-11}
Code Version                 &&&&  \cc{c}{CUDA}     & \cc{c}{OACC}  &&&& \cc{c}{CUDA}	  & \cc{c}{OACC}\\
\midrule
$T_{\text{Prop}}$ [msec]     &&&&     13.78	    &  13.91	    &&&&  7.60  	  &  7.51  \\
GB/s                         &&&& 	   169	    &	 167	      &&&&  306		    &  310   \\  
${\cal E}_p$                 &&&&      59\%	    &	58\%	      &&&&  64\%  	  &  65\%  \\	
\midrule                
$T_{\text{Bc}}$ [msec]       &&&&      4.42	    &  2.76	      &&&&  1.11  	  &  0.71  \\  
\midrule           
$T_{\text{Collide}}$ [msec]  &&&&     39.86	    &  78.65	    &&&&  16.80 	  &  36.39 \\
MLUPS                        &&&& 	     99	    &	  50	      &&&&  234		    &  108   \\
${\cal E}_c$                 &&&&      45\%	    &	23\%	      &&&&  52\%  	  &  24\%  \\  
\midrule
$T_{\text{WC}}$/iter [msec]  &&&&     58.07	    &  96.57	    &&&&  26.84 	  &  44.61 \\ 
MLUPS                        &&&&     68	      &  41	        &&&&  147		    &  88    \\ 
\bottomrule
\end{tabular}
\end{table}
%

%
\section{GPU Results}\label{sec:results}

We start our performance analysis on NVIDIA GPUs, comparing our OpenACC code 
with an implementation of the same algorithm written in CUDA~\cite{caf11,sbac-pad13} 
and optimized for Fermi and Kepler architectures. In other words, we compare 
with a low-level programming approach which gives programmers more freedom in 
mapping codes and data on GPU architectures, and then more optimization options.

Table~\ref{comparison} summarizes performance figures of codes on a reference 
lattice of $1920 \times 2048$ sites run on two NVIDIA systems, the K40 and K80 
boards. 
These accelerator cards are powered by respectively the GK110B and GK210 
processors based on the latest {\em Kepler} GPU family.
The K40 has a peak memory bandwidth of 288 GB/s, and a peak floating-point 
performance of 1430 GFLOPs; this can increase up to 
1660 GFLOPs boosting the GPU clock frequency to 875 MHz.
The NVIDIA K80 is a dual-GPU system. Each GPU features a peak memory 
bandwidth of 240 GB/s, and a peak double-precision floating-point performance 
of 935 GFLOPs; this can be increased up to 1455 GFLOPs, again
boosting the GPU clock to 875 MHz. 

We have used the PGI compiler version 14.10 for our test on the K40 and 
version 15.10 for running on the K80; while for CUDA we have used the 
NVIDIA compiler version 6.5 for the K40 and version 7.5 for the K80. Using more 
recent version of the compilers for the K40 does not changes the results.
The codes executed on the K80 system runs two MPI ranks, each using one 
GPU of the same accelerator card.

%
\begin{figure}
\centering
\includegraphics[width=\textwidth]{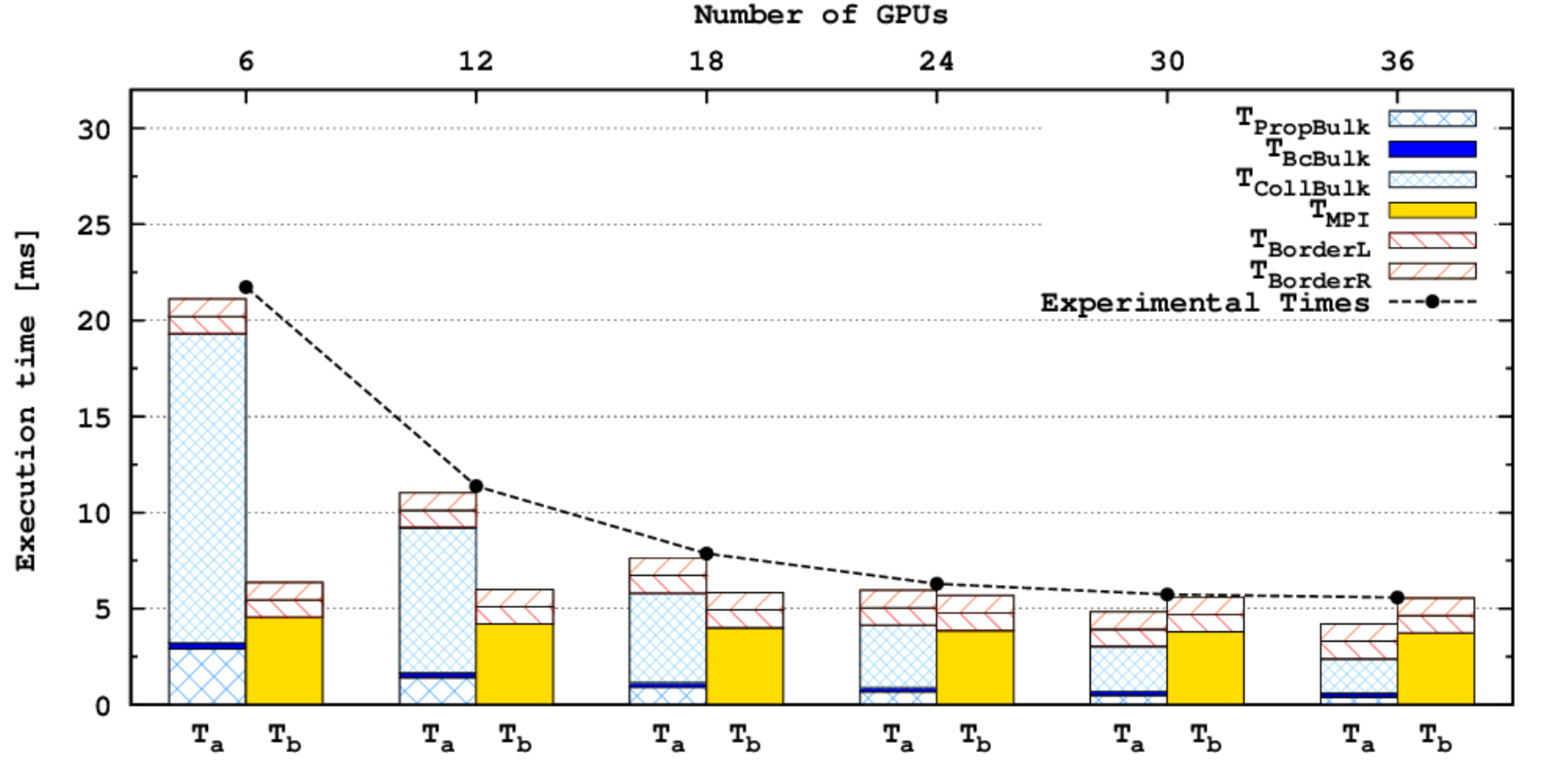}
\caption{\label{fig:histo} $T_a$ and $T_b$ for the time model 
defined in the text on a lattice of $1080 \times 5736$ points
as a function of the number of GPUs. 
The black points are the execution times of the code 
with all asynchronous steps enabled.
}
\end{figure}
%

The first line of \tablename~\ref{comparison} refers to the execution of 
{\tt propagate} kernel.
We show the execution time, the effective bandwidth, and the efficiency 
${\cal E}_p$ computed w.r.t. the peak memory bandwidth of each system; 
the table then lists execution times of the {\tt bc} function, showing 
that this routine has limited (albeit non negligible) impact on performance.
For the {\tt collide} kernel, we show the execution time and the efficiency 
${\cal E}_c$ as a fraction of peak performance.
Efficiency is computed using as number of double-precision operations 
for each lattice-site either the number measured by the profiler through 
the hardware counters available on the processor or 
the number of floating-point instructions of the corresponding assembly code. 
Finally, the last two lines at bottom show the wall-clock execution time (WcT) 
and the corresponding {\em Millions Lattice UPdate per Second} (MLUPS) 
-- counting the number of sites handled per second -- of the full 
production-ready code.

For {\tt propagate}, which is strongly memory bound, the CUDA and OpenACC 
versions run at $\approx 60\%$ of peak. 
For the {\tt collide} kernel, which is the most computationally intensive part 
of the application, the OpenACC code has an efficiency of $\approx 25\%$ on 
each system, while the CUDA version doubles this figure, running at $45\%$ 
of peak on the K40 and $52\%$ on the K80 thanks to the higher number of 
available registers.

%
\begin{figure}[t]
\centering
\begin{lstlisting}[language={},basicstyle=\footnotesize]
DFMA R84, R78, c[0x3][0x250], RZ;
LDG.E.64 R36, [R4];
TEXDEPBAR 0x1e;
DADD R4, R6, R12;
DFMA R6, R16, c[0x3][0x130], R82;
DFMA R80, R16, c[0x3][0x258], R84;
TEXDEPBAR 0x1d;
\end{lstlisting}
\caption{\label{fig:collide_sass} Part of SASS assembly of the CUDA collide 
kernel showing the use of constant cache values (operands {\tt c[...][...]}) 
addressed directly by instructions. 
}
\end{figure}
%

Our analysis of the performance gap between OpenACC and CUDA codes for the 
{\tt collide} kernel shows that a crucial role is played by the different way 
in which the coefficients of the Hermite polynomial expansion are stored.
Our code uses 18 double-precision Hermite coefficients for each population, and 
the associated memory footprint to store them is 2368 $( = 18 \times 37 \times 8 )~\mbox{Bytes}$.
Coefficients are initialized at run-time by the host and used as constant values 
by the kernels running on the GPU.
CUDA allows explicit control on the allocation of data onto the various memory structures 
inside the GPU; in our case, the 64~KB low-latency constant cache of each GPU processor-core (SMX) is large enough to fit all coefficients, 
so they are copied there once before starting execution of the main loop. 
Data items stored in constant cache can be directly addressed by assembly instructions and 
do not require load operations onto registers (see figure \ref{fig:collide_sass}). As a consequence, a larger number of general 
registers is then available to the compiler to fully unroll all inner loops of the {\tt collide} 
routine.
This allows to cache the accesses to the populations of a site onto registers improving 
performance significantly.
Note also that data stored on the constant-memory is available to all threads, so no data 
replication on general registers is needed, and the performance impact is relevant in spite 
of the relatively small number of coefficients.

At a variance with the CUDA case, the OpenACC compiler can not identify the Hermite 
coefficients as constant values as they are initialized at run-time.
The consequences of this are that: (a) they are loaded on the global memory, and accesses 
to them are handled as less efficient regular memory accesses; 
(b) registers are required to stage these coefficients, and for this reason inner loops 
can not be unrolled and code runs approximately 2X slower w.r.t. the CUDA version. 
Unrolling the inner-loops of collide by hand causes significant register-spilling 
(5432 bytes spill stores, 13368 bytes spill loads) harming the performance by approximately 
a factor 10X w.r.t the CUDA version. 
As a check of this analysis, we have verified that a CUDA version that does not use the 
constant-memory and does not unroll inner loops matches the performance of the OpenACC 
code version.

Drawing a temporary conclusion looking at the overall MLUPS delivered of the full 
code on each system, we have that performances of OpenACC code are $\approx 40\%$ 
lower with respect to the CUDA code.



%
\begin{figure}
\centering
\includegraphics[width=\textwidth]{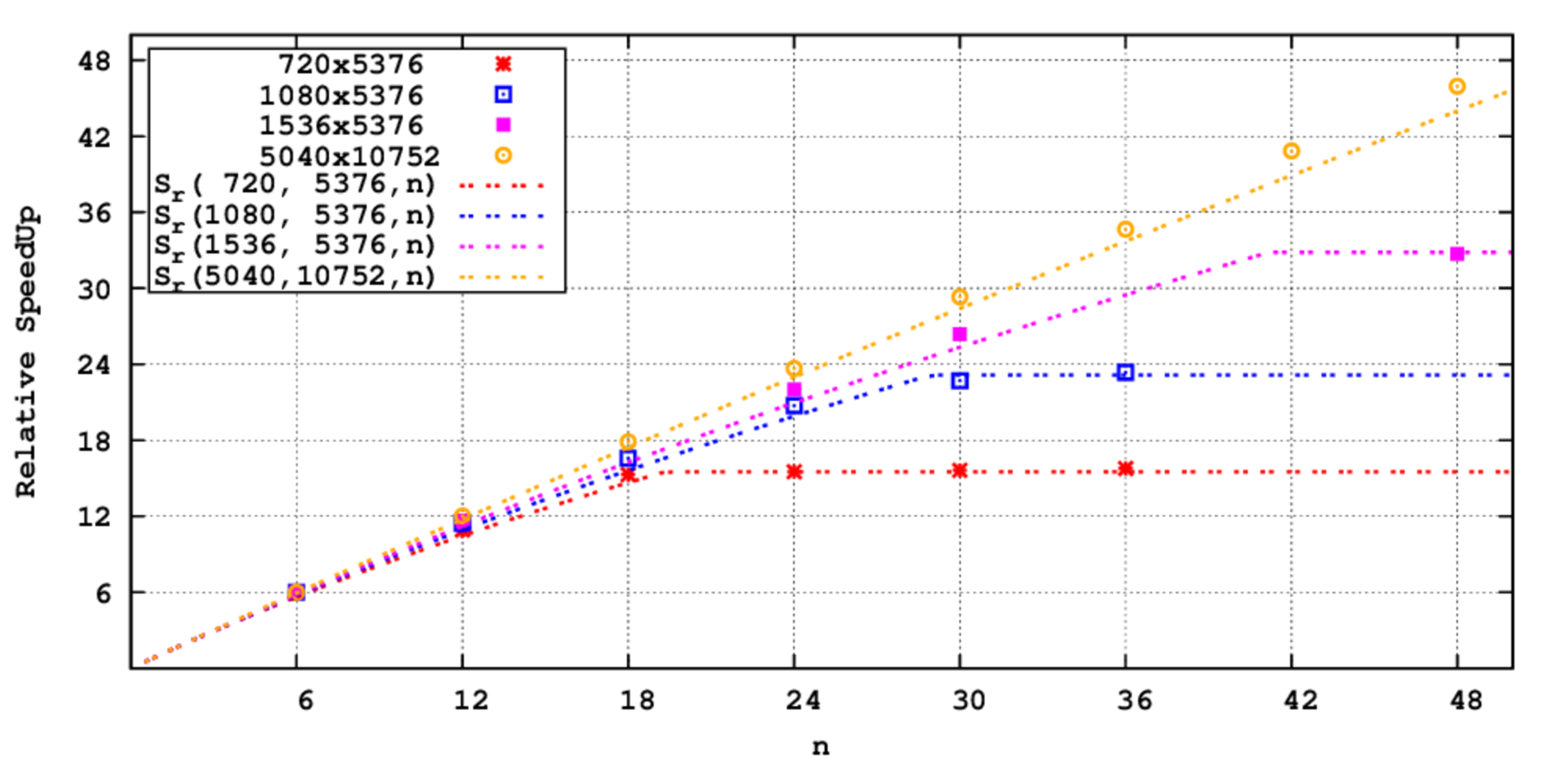}
\caption{\label{fig:scalability} Strong scaling behavior of the OpenACC code 
as a function of the number of GPUs ($n$) for several lattice sizes. 
Points are experimental data and dashed lines are the predictions of 
our timing model.}
\end{figure}
%

%
\begin{figure}
\centering
\includegraphics[width=\textwidth]{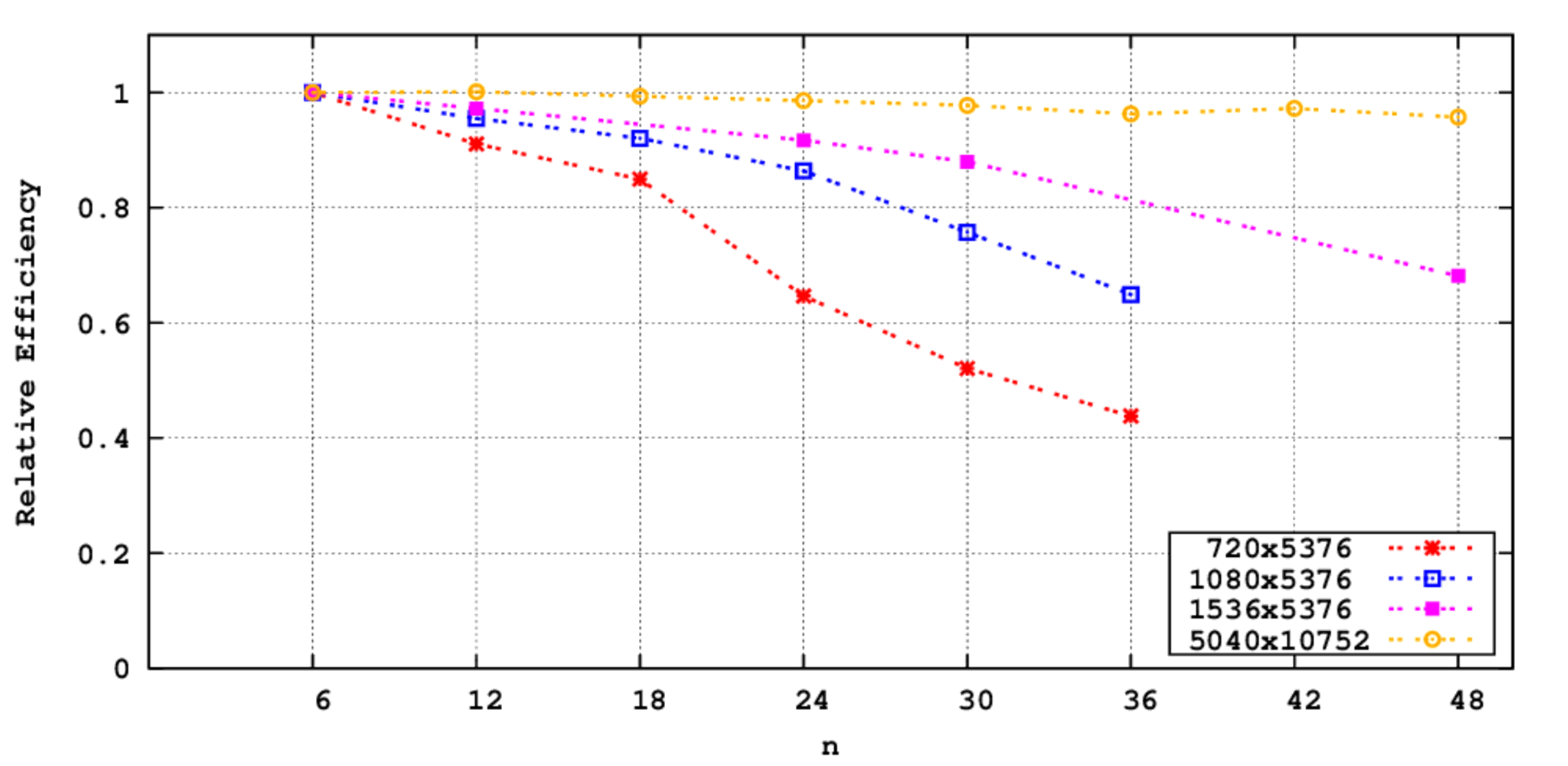}
\caption{\label{fig:efficiency} Parallel efficiency of the OpenACC code 
as a function of the number of GPUs ($n$) for several lattice sizes.
}
\end{figure}
%

%
We now discuss in details the scaling behavior of our parallel implementation 
as we run it on an increasing number of GPUs.
We model the execution time of the whole program as $T \approx \max\{T_a, T_b\}$, 
with $T_a$ and $T_b$ defined as:
\begin{equation*}
T_a = T_{\text{bulk}} + T_{\text{borderL}} + T_{\text{borderR}}~,~~~~~~~~~
T_b = T_{\text{MPI}}  + T_{\text{borderL}} + T_{\text{borderR}}
\end{equation*}
%
%
and $T_{\text{bulk}}, T_{\text{borderL}}, T_{\text{borderR}}$ 
are respectively the sums of the execution times of {\tt propagate}, 
{\tt bc} and {\tt collide} on the bulk, and on the left and right halos, 
while $T_{\text{MPI}}$ refers to MPI communications;
we first profile the execution time of each kernel and MPI communication 
running them in sequence, i.e. without any overlap, and then we measure the 
execution time of the whole program with all asynchronous steps enabled.

This model is in good agreement with data measured on an
Infiniband-interconnected  cluster with 36 GPUs (6 GPUs on each node): 
\figurename~\ref{fig:histo} shows our measured data for $T_a$ and $T_b$ on a 
lattice of $1080 \times 5736$ points. The histograms show 
the times taken by each section of the code when running serially while
the black dots show the time taken by the asynchronous code.   
For this choice of the lattice size, we see that $T \approx T_a$ up to 
24 GPUs as communications are fully hidden behind the
execution of the program on the bulk; as long as this condition holds, the
code enjoys full scalability. 
As we increase the number of GPUs ($\ge 30$) $T \approx T_b$, 
communications become the bottleneck and the scaling behavior necessarily degrades.

%
%

We further characterize the execution time assuming, to first approximation,   
that bulk processing is proportional to $(L_x \times L_y)$, boundary conditions scale 
as $L_x$, and communication and border processing scales as $L_y$; so, on $n$ GPUs
%
\[
 T(L_x, L_y, n) = \max \left\{ \alpha \frac{L_x}{n} L_y + \beta\frac{L_x}{n}, \:\:\:\: \gamma L_y \right\} + \delta L_y
\]
%
We extract the parameters ($\alpha$, $\beta$, $\gamma$ and $\delta$) from the 
profiling data of \figurename~\ref{fig:histo}, and define the function  
\begin{equation*}
  S_r(L_x,L_y,n) = \frac{T(L_x, L_y, 1)}{T(L_x, L_y, n)}
\end{equation*}
to predict the relative speedup for any number of GPUs and any lattice size. 
\figurename~\ref{fig:scalability} shows the (strong) scaling behavior of 
our code for several lattice sizes relevant for physics simulations; 
dots are measured values and dashed lines are plots of $S_r()$ 
for different values of $L_x$ and $L_y$.  
Values of $S_r()$ are in good agreement with experimental 
data, and predict the number of GPUs at which scaling violations start to become important.  
For large lattices ($5040 \times 10752$) the code has an excellent 
scaling behavior up to 48 GPUs, slightly underestimated by our model  
as constants are calibrated on smaller lattices so they are more sensitive 
to overheads.
\figurename~\ref{fig:efficiency} shows the corresponding parallel efficiency 
of our code. For large lattices it remains close to one, while for smaller 
lattice using the largest number of GPUs is in the range 40-60\%.

%
\begin{table}[t]
\centering
\caption{Selected hardware features of the computing systems considered 
in this paper to assess portability of OpenACC codes. Clock freq. for Intel Xeon E5-2680 v3 are for AVX code.}
\label{tab:architectures}
\begin{tabular}{l r rl r r r r}
\toprule 
                               &  Intel Xeon      & \mc{2}{c}{NVIDIA K80} & AMD S9150 \\  
\midrule
processor codename             &  E5-2630 v3      & \mc{2}{c}{GK210}      & Hawaii XT \\
\#physical-cores               &  8               & 13    &   x 2         & 44        \\
\#logical-cores                &  16              & 2496  &   x 2         & 2816      \\  
nominal clock Freq. (GHz)      &  2.1             & 0.562 &               & 0.900     \\
Nominal GFLOPS (DP)            &  268.625         & 935   &   x 2         & 2530      \\
Max Boosted clock Freq. (GHz)  &  2.6             & 0.875 &               & --        \\
Boosted GFLOPS (DP)            &  331.56          & 1455  &   x 2         & --        \\ 
Max Memory (GB)                &  768             & 12    &   x 2         & 16        \\
Mem Bandwidth (GB/s)           &  59              & 240   &   x 2         & 320       \\
ECC                            & YES              & YES   &               & YES       \\
\bottomrule
\end{tabular}
\end{table}
%


\section{OpenACC Portability}\label{sec:portability}

In this section we analyze the portability of the OpenACC version described in
the previous sections across several architecture taking into account
also portability of performances.

In this work, we consider three different target computing systems
supported by the PGI OpenACC compiler: an {\em x}86 commodity multi-core
processor, and two accelerators, an NVIDIA K80 and an AMD FirePro S9150
GPU.
Together with the Intel Xeon-Phi this set of architectures is currently 
the most used in HPC systems. The PGI compiler so far does not support Xeon-Phi
so we cannot include this accelerator in our comparison; for performance 
details of our LB code on Xeon-Phi using other frameworks see~\cite{iccs13,europar14}. 
 
The Intel Xeon processor is an 8-core E5-2630 v3 CPU based 
on the recent Haswell micro-architecture. For this processors the 
base clock frequency is 2.1 GHz for applications using the AVX vector-unit, like the LB code analyzed here; enabling turbo-mode and 
using all 8 cores available the frequency can boosts up to 2.6 GHz for these kind of applications. This CPU has then 
a peak memory bandwidth of 59 GB/s, and a a peak double-precision floating-point 
performance of $\approx 330$ GFLOPs; the PGI compiler supports generating multi threaded code from OpenACC for this family of processors since version 15.9. 
%
%
The NVIDIA K80 system (whose details have been introduced in Sect.~\ref{sec:results}) 
is a dual-GPU system, and for our benchmarks we have used only one of the two GK210 
GPUs, whose features are very similar to those of the NVIDIA K40 GPU considered in 
previous sections, so our results can be directly compared with those shown in
\tablename~\ref{comparison}. 
Finally, the AMD S9150 is a GPU accelerator with a memory bandwidth of 320 GB/s 
and  up to 2620 GFLOPs double-precision floating point performance.
Table~\ref{tab:architectures} summarizes several hardware 
parameters useful for our analysis.

The target architecture for the compilation is specified by appropriate
options  (e.g., {\tt -ta=radeon} and {\tt -ta=nvidia} for AMD and NVDIA
GPUs respectively and {\tt -ta=multicore}  for {\em x}86 multicore CPUs).

%
\begin{figure}
\centering
\begin{lstlisting}[language=C,basicstyle=\footnotesize]
#pragma acc kernels present(prv) present(nxt)
#pragma acc loop gang independent
for ( ix=HX; ix < (HX+SIZEX); ix++) {
  site_i = (ix*NY);
  
  #pragma acc loop seq independent
  for ( ipop=0; ipop<37; ipop++ ) {
  
    #pragma acc loop vector independent
    for ( iy=HY; iy < (HY+SIZEY); iy++) {
      nxt[ipop*NX*NY+site_i+iy] = prv[ipop*NX*NY+site_i+iy+OFFSET[ipop]];
    }
     
  }
}
\end{lstlisting}
\caption{\label{fig:propagate2} Code snapshot of the {\tt propagate} kernel, 
version v2, that improves performances on {\em x}86 CPUs.}
\end{figure}
%

In the {\em x}86 case, the parallelization performed by the compiler is similar to that
implied by the {\tt omp parallel} OpenMP directive. 
Gangs of OpenACC loops and regions are run on different physical-cores,
or  virtual-cores if hyper-threading is enabled. The compiler  uses all
available cores  on the processor unless a different number is
specified by the {\tt gang} clause or through the {\tt ACC\_NUM\_CORES}
environment variables. The directive {\tt acc vector} is considered as a hint
indicating that the compiler can vectorize the loop, but the
compiler uses its own analysis to determine if vectorization 
can be applied, and generate SIMD code.  A multicore CPU is treated as
a shared-memory accelerator, so data clauses  (like {\tt copy, copyin,
copyout, create}) are ignored and no data copies are executed. 

In all cases, it is useful to  keep an eye on the strategies taken by 
the compiler using  compiler options {\tt -Minfo} or {\tt -Minfo=accel} 
that enable feedback messages, giving some details on the parallel 
and/or vectorized code generated.

We have run and benchmarked the {\tt propagate} and {\tt collide}
kernels  on the three systems described above. From the point of view of code
portability results are really good as exactly the same C-code (the one
described in previous sections), annotated  with the same OpenACC
pragmas, immediately runs on all three architectures.

%
\begin{table}[b]
\centering
\caption{Benchmark results for the {\tt propagate} and {\tt collide} kernels 
on three different processor architectures. Both kernels are compiled 
using the PGI compiler, version 15.10.}
\label{tab:benchmark}
\begin{tabular}{l c c c}
\toprule
                       & E5-2630 v3 & GK210      & Hawaii XT  \\
\midrule
propagate v1 [GB/s]    &    10      & 155        & 216        \\
propagate v2 [GB/s]    &    32      & 145        & 223        \\
\midrule
collide v1 [MLUPS]     &     8      & 55         & 53         \\
collide v2 [MLUPS]     &    12      &  6         &  5         \\
\bottomrule
\end{tabular}
\end{table}
%
  
However results are not fully satisfying from the point of view of performance 
portability, see table~\ref{tab:benchmark}, codes version v1. 
For instance, performances on the AMD GPU increases for {\tt propagate} 
in line with the higher available memory bandwidth but performance for 
{\tt collide} remains roughly the same for both GPUs, in spite of the 
significantly higher peak value of the AMD accelerator. 
On the {\em x}86 CPU, results are even more worrying, with a larger 
drop in performance for both kernels. 

Trying to improve performances, we experimented with a different organization 
of the {\tt propagate} kernel: while in the original code 
(version v1, see again \figurename~\ref{fig:propagate})  
for each lattice site we move all 37 populations associated to it, 
the new version (called v2, see \figurename~\ref{fig:propagate2}) 
processes lattice-sites by columns: for each column and for each population 
index we handle in sequence all sites of the column. 

Performances of the two kernels run on all target architectures 
are shown on table \tablename~\ref{tab:benchmark}. 
Comparing version v2 with v1 we see that on CPUs it gives better performances  
increasing the effective memory bandwidth by a factor 3X. The measured value is 
32 GB/s corresponding to 54\% of the raw peak. On NVIDIA and AMD GPUs both 
versions give approximately (within 10\%) the same result. 

We have also experimented with a new version of the collide kernel 
(version v2) in which we unroll most internals loops by hand. Also 
in this case version v2 gives better results on {\em x}86 CPUs 
increasing performance by a factor 1.5X w.r.t. version v1. 
On GPUs however version v2 is much slower. Unrolling most internal loops increases the
number of registers needed per thread. This causes register spilling and the resulting
local memory overhead causes the slowdown of v2 on GPUs.
As shown in \figurename~\ref{fig:unrollclause} a {\tt unroll} clause for the {\tt loop}
directive combined with the {\tt device\_type} clause would allow to maintain the same code
for multicore CPUs and GPUs.

%
\begin{figure}
\centering
\begin{lstlisting}[language=C,basicstyle=\footnotesize]
#pragma acc kernels present(prv) present(nxt) present(param[0:1])
#pragma acc loop gang independent
for ( ix = HX; ix < (HX+SIZEX); ix++) {

  #pragma acc loop vector(NVECTOR_C) independent
  for ( iy = HY; iy < (HY+SIZEY); iy++) { 
    ...
    #pragma acc loop seq device_type(multicore) unroll
    for( i = 0; i < NPOP; i++ ) {
      ...
    }    
    ...
  }
  
}
\end{lstlisting}
\caption{\label{fig:unrollclause} Example usage of the proposed {\tt unroll} clause.}
\end{figure}
%

\tablename~\ref{comparison2} collects our final comparison results; it 
summarizes the best results that we have measured on all target architectures 
using the OpenACC PGI compiler and the performances of the same computational kernels 
coded with other programming approaches, closer to each specific architecture.  
In details, for {\em x}86 CPUs we show results of two multi-thread codes 
compiled with the Intel compiler version 15: one uses intrinsics functions~\cite{ccp12,caf13} 
to exploit vectorization, while the latter~\cite{ppam15} uses OpenMP directives; both 
use OpenMP to handle multicore parallelization.
For NVIDIA GPUs we consider the CUDA code, while for AMD GPUs we have used 
GCC and OpenCL~\cite{iccs14,europar14}. 


\begin{table}
\caption{
Performance comparison of various programming frameworks on various processors.
MLUPS stands for Mega-Lattice Updates per Second; the performance in GFLOPs is
obtained assuming that each lattice points uses 6500 floating-point operations; 
Total performance (Tot perf.) is evaluated on a program that invokes in 
sequence {\tt propagate} and {\tt collide}.}
\label{comparison2}
\resizebox{\textwidth}{!}{
\begin{tabular}{ l l  rrr | r rr | r rr }
\toprule
                             && \mc{3}{c|}{E5-2630 v3}  	         && \mc{2}{c|}{GK210}	      && \mc{2}{c}{Hawaii XT}	   \\
\midrule
compiler                     && ICC 15     & ICC 15	 & PGI 15.10    && NVCC 7.5	& PGI 15.10    && GCC	     & PGI 15.10  \\
model                        && Intrinsics & OMP	   & OACC         && CUDA  	  & OACC         && OCL	     & OACC	 \\
\midrule  
%
{\tt propagate} perf. [GB/s] && 38	       & 32 	   & 32	          && 154		  & 155	      && 232	     & 216  \\
${\cal E}_p$                 && 65\%	     & 54\%	   & 54\%         && 64\%  	  & 65\%      && 73\%	     & 70\% \\
%
\midrule
%
{\tt collide} perf. [MLUPS]  && 14	     & 11 	   & 12             && 117		  & 55	        && 76	       & 54    \\
{\tt collide} perf. [GFLOPs] && 92	     & 71 	   & 78	            && 760		  & 356	        && 494	     & 351   \\ 
${\cal E}_c$                 && 28\%	   & 22\%	   & 24\%           && 52\%  	  & 24\%        && 19\%	     & 14\%  \\
\midrule
Tot perf. [MLUPS]            && 11.5	   & 9.2	 & 9.8         && 80.7  	& 45.6        && 63.7	     & 47.0  \\
\bottomrule
%
%
\end{tabular}
}
\end{table}


\tablename~\ref{comparison2} provides several metrics for a comparative 
assessment of code portability: for compute-intensive kernels  
one may consider the ratio of delivered floating-point performance w.r.t 
peak performance on different machines; 
for data-intensive application a better parameter may be the ratio of measured 
memory bandwidth w.r.t peak; 
finally one may want to compare the performance of the OpenACC code with the 
corresponding performance of the same code written in an architecture-specific 
programming language. 

Having this in mind, several comments are in order:
\bi
\item 
on the {\em x}86 CPU, the three versions of the code have remarkably the 
same level of performances for both kernels. All runs are using all 8 CPU-cores available and have turbo-mode enabled. Using intrinsics is slightly 
more efficient, but the more ``user-friendly'' programming frameworks are 
almost as efficient (within $\approx 10\%$). For the collide 
kernel the main bottleneck for performances is the memory latency in accessing 
Hermite coefficients and population values needed to compute this kernel 
for each lattice site.
\item 
On the NVIDIA GPU, the OpenACC code gives the same performance of 
CUDA for the {\tt propagate} kernel. 
Results are not the same for the {\tt collide} kernel where we measure a drop of 2X 
compared w.r.t the CUDA version. Reasons for this behaviour have been 
analyzed in the Sect.~\ref{sec:results}.
\item
On the AMD GPU, we have almost equally efficient {\tt propagate} versions 
with the two available programming environments; but the performance of the {\tt collide}
kernel is lower for OpenCL and OpenACC. In this case, we have not analyzed 
the performance bottleneck in depth; 
for this reason it is possible that performance could be optimized further.
%
\ei
 
Trying to provide a global assessment of the performance portability offered 
by the PGI OpenACC compiler, we first remark that it is able to produce very 
efficient code for the {\tt propagate} kernel exploiting a large fraction of 
the memory bandwidth offered by all processors; this is substantiated by the 
reported figures for ${\cal E}_p$ (i.e., by the sustained memory bandwidth 
w.r.t. peak memory bandwidth).

For the compute intensive collide kernels, OpenACC comes at the cost of a non 
negligible performance gap; considering in this case ${\cal E}_c$ 
(measuring the actual floating point performance), we see a performance drop of 2X.

As a global assessment, the last line of \tablename~\ref{comparison2} shows 
an effective performance metric for the whole code assuming to execute  
{\tt propagate} and {\tt collide} in sequence; this gives a lower-bound of  
performances of the program since the two routines can often (actually 
depends on routines to be run between the two) be merged in one single step.
Keeping this point in mind, our figures show that OpenACC is indeed able to support 
code portability at the price of a performance drop lower than $50\%$, and 
with reasonable expectations of further improvements in the future.
As an unexpected aside, we also show that the PGI compiler is  
remarkably efficient for multi-core {\em x}86 CPUs.

%
\begin{figure}
\centering
\includegraphics[width=\textwidth]{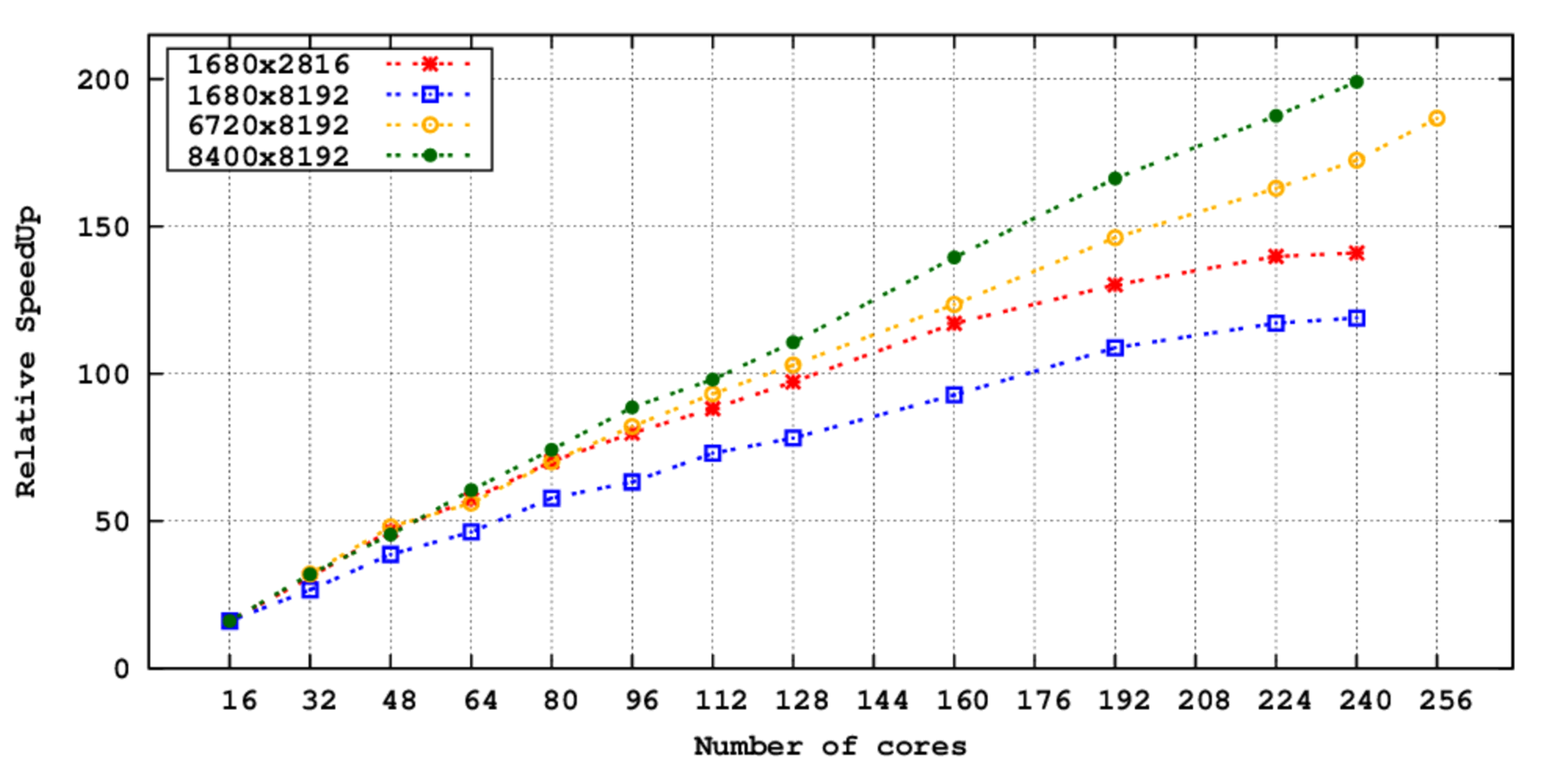}
\caption{\label{galileo-speedup} Strong scaling behaviour of our OpenACC code 
as function of the number of CPU-cores. The code run on a {\em X}86 CPU cluster 
interconnected via Infiniband QDR network.}
\end{figure}
%

\begin{figure}[b]
\centering
\includegraphics[width=\textwidth]{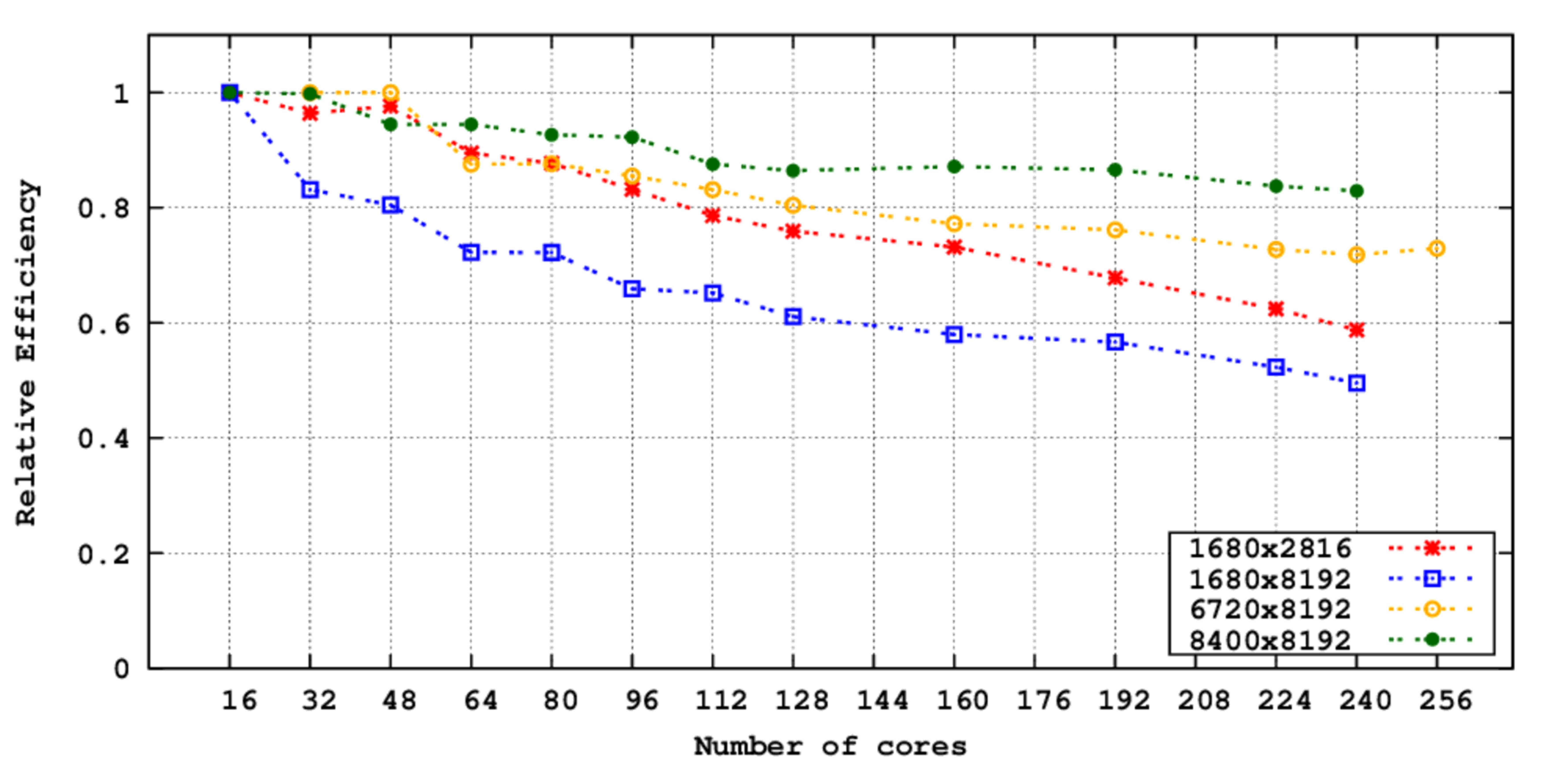}
\caption{\label{galileo-efficiency} Parallel efficiency of our OpenACC code 
as function of the number of CPU-cores. The code run on a {\em X}86 CPU cluster 
interconnected via Infiniband QDR network.}
\end{figure}

Finally, we have successfully run the same OpenACC code used for the GPU cluster 
also on a cluster of Intel Xeon E5-2630 v3 CPUs.
Each node of the cluster hosts two eight-core CPUs, and the nodes are 
interconnected via Infiniband QDR network.
The code runs two MPI ranks per node, and each MPI process ``offloads'' 
execution of the kernels on the 8-cores of the CPU on which it is running.
\figurename~\ref{galileo-speedup} shows the strong scalability behaviour 
achieved on 16 nodes of the cluster corresponding to 32 CPUs and 256 cores, 
using several lattice sizes with different aspect ratios. \figurename~\ref{galileo-efficiency} 
shows the corresponding parallel efficiency. 
Contrary to the case of a GPU cluster, in this case we do not have a 
real overlap between communication and computation. Both are concurrently 
managed by the cores of CPUs, and this limits scaling behaviour. 
For the lattices we have tried, using the largest number (240-256) of 
CPU-cores the relative speedup is in the range 120-200, and the parallel efficiency 
results in the range 50-80\%.


%
%


\section{Conclusions}\label{sec:conclusions}

In conclusion, we have successfully ported, tested and benchmarked a complete 
multi-node Lattice Boltzmann code using OpenACC, and characterized its 
performances through an accurate timing model.

Our experience with OpenACC is very positive from the point of view of 
code portability and programmability.
The effort to port existing codes to OpenACC is reasonably limited and easy 
to handle; we started from an existing {\tt C} version 
and marked through {\em directives} regions of code to offload and 
run on accelerators, instructing the compiler to identify and exploit 
available parallelism.
However, we have to underline that some major changes in the global structure 
of codes and in data organization cannot be handled automatically by compilers 
and the overall organization must be (at least partially) aware of the target 
architectures. 
For example, in our case it is crucial to organize data as {\em Structure of Arrays} 
to allow to coalesce performance-critical memory accesses, and exploit vectorization.

Concerning code portability across different target architectures, our 
experience shows that the PGI compiler easily allows to target different processors 
widely used today in most HPC systems. 
As we have shown in Sect.\ref{sec:portability} the latest version of the compiler 
is able to target both {\em x}86 multi-core CPUs and NVIDIA and AMD GPUs. 
We consider this a major result enabling users to quickly and easily benchmark a
single-code  on a wide range of target processors and decide which hardware
better fits the computing  requirements of applications.
However also in this case we have to say that for optimal performances some
changes in the organization of the codes may be required. 
In our case, for example, on {\em x}86 multi-core CPUs we improved the
performance of the {\tt propagate}  kernel by a factor 3X changing the order of
processing of the lattice sites.
On the other hand, applying this changes on codes written in a high level language 
annotated by OpenACC is much simpler than doing the same on codes
written in OpenCL or heavily using intrinsic functions.
This of course improves productivity of programmers and encourages 
experimentation with different codes.
 
Concerning performance results, one is ready to accept that using a high  
level programming model trades better programmability with computing 
efficiency, and a performance drop $\le 20\%$ may be considered a 
satisfactory result. 
Our experiments show that actual performances drop can be larger, 
approximately around a 2X factor in our case, with overall performances 
that may be $\approx 50$\% of what is made possible by using more 
processor-specific programming methodologies. 
In many cases, as explained and discussed in Sect.~\ref{sec:results} 
we understand the reasons behind this gap and have good reasons to believe 
that future versions of the compiler may introduce supports to narrow 
this performance gap. 
For example, in our case support for the constant cache available on NVIDIA
GPUs would be useful.
As an interim step, one may work around, e.g. exploiting 
the interoperability between OpenACC and CUDA for NVIDIA GPUS,  
to foster the high productivity of OpenACC and still get full 
performance by using CUDA for the most performance critical kernels. 
Similar approaches also apply to CPUs and AMD GPUs.

We believe that our analysis provides important feedback to help users 
understand the capabilities of the OpenACC approach as well as several hints 
to improve the performance of OpenACC codes. 

In the short term future, we plan to test our OpenACC LB codes on yet 
more processors architectures, e.g. Intel Xeon-Phi, as soon as OpenACC 
support becomes available. 


\section*{Acknowledgments.}

This work was done in the framework of the COKA, COSA and Suma projects of INFN.  
We thank INFN-Pisa (Pisa, Italy), the NVIDIA J\"ulich Application Lab (J\"ulich
Supercomputer Center, J\"ulich, Germany) and the CINECA (Bologna, Italy) 
for allowing us to use their computing systems.
AG has been supported by the European Union's Horizon 2020 research and
innovation programme under the Marie Sklodowska-Curie grant agreement No 642069.






\end{document}